\documentclass{ws-mpla}
\usepackage[super]{cite}
\usepackage{xspace}
\usepackage{hyperref}
\hypersetup{colorlinks,urlcolor=black,citecolor=black,linkcolor=black,filecolor=black}
\usepackage{graphicx}
\usepackage{breakurl}
\begin{document}
\markboth{W. Barter}
{Measurements of Electroweak Bosons at LHCb}

%%%%%%%%%%%%%%%%%%%%% Publisher's Area please ignore %%%%%%%%%%%%%%
\catchline{}{}{}{}{}
%%%%%%%%%%%%%%%%%%%%%%%%%%%%%%%%%%%%%%%%%%%%%%%%%%%%%%%%%%%%%%%%%%%

\title{A BRIEF REVIEW OF MEASUREMENTS OF ELECTROWEAK BOSONS AT THE LHCB EXPERIMENT IN LHC RUN 1
}

\author{\footnotesize WILLIAM BARTER}

\address{European Organisation for Nuclear Research, Geneva, Switzerland\\
william.barter@cern.ch}

\newcommand{\sttw}{\ensuremath{\sin^2\theta^\text{lept.}_\text{eff.}}\xspace}
\newcommand{\phist}{\ensuremath{\phi^{*}_\eta}\xspace}
\def\rd{{\rm{d}}}

\maketitle

\pub{Received (Day Month Year)}{Revised (Day Month Year)}

\begin{abstract}
The LHCb experiment is one of four major experiments at the LHC. Despite being designed for the study of beauty and charm particles, it has made important contributions in other areas, such as the production and decay of $W$ and $Z$ bosons. Such measurements can be used to study and constrain parton distribution functions, as well as to test perturbative quantum chromodynamics in hard scattering processes. The angular structure of $Z$ boson decays to leptons can also be studied and used to measure the weak mixing angle. The phase space probed by LHCb is particularly sensitive to this quantity, and the LHCb measurement using the dimuon final state is currently the most precise determination of $\sin^2\theta^\text{lept.}_\text{eff.}$ at the LHC. LHCb measurements made using data collected during the first period of LHC operations (LHC Run 1) are discussed in this review. The article also considers the potential impact of related future measurements.
\end{abstract}

\ccode{PACS Nos.: include PACS Nos.}

\section{Introduction}
Electroweak bosons exhibit a rich phenomenology that make them worthy of study at the LHC.\footnote{This article will denote $Z/\gamma^*$ production as $Z$ boson production for simplicity.} The factorization theorem states that the cross-section for electroweak boson production in hadron collisions can be calculated from the convolution of a partonic cross-section, calculable with perturbative methods, and parton distribution functions (PDFs) that describe the partonic structure of the colliding hadrons, that are determined by non-perturbative long-distance interactions within quantum chromodynamics (QCD).\cite{Collins:1989gx} The partonic cross-section is sensitive both to electroweak theory, and to corrections that arise within QCD. Integrating over all possible initial state configurations gives the cross-section for hadronic, as opposed to parton-level, collisions. This means that measurements of hard processes, such as electroweak boson production, are able to probe both PDFs and the hard partonic interaction; different differential distributions provide sensitivity to different phenomenological aspects of the Standard Model.

It is perhaps best to split discussion of these studies in two. This article first considers measurements of cross-sections and related quantities at LHCb, which are generally used to probe and test QCD, either within the proton, studying PDFs, or within the hard interaction itself, where the modeling of higher order effects in perturbative QCD (pQCD), such as parton emission, can be tested by studying the transverse momentum of electroweak bosons and related variables. This article then considers a precision test of electroweak theory at LHCb: the measurement of the weak mixing angle. At each stage this article will set out the key motivations for performing such measurements in the forward rapidity region probed by LHCb. The measurements and results discussed here are based on data collected in the first period of LHC operations (LHC Run 1), where protons were collided at $\sqrt{s} = 7$ and $8$~TeV, though the potential for future studies is also outlined. However, before specific measurements and their motivation are discussed, this review begins with a brief summary of the LHCb detector.

\section{The LHCb detector at the LHC}
The LHCb experiment at the LHC was designed for the study of beauty and charm particles. It is a single arm spectrometer providing dedicated coverage close to the beamline where a large fraction of heavy flavor particles are produced.  Uniquely at the LHC, the LHCb detector provides precision coverage of this forward region (covering a pseudorapidity range $2<\eta<5$), occupying a complementary phase-space to ATLAS and CMS, which provide precision coverage of the central region (different sub-detectors at these experiments offer different angular coverage, though tracking systems extend to $|\eta| \sim 2.5$, and both ATLAS and CMS have forward calorimeter coverage). The key features of the LHCb experiment are set out in Ref.~\refcite{lhcb-det}, though it is worth highlighting a few important features of the detector and detector operations, and differences with respect to the other LHC detectors, relevant for studying the physics associated with electroweak bosons.
\begin{itemize}
\item LHCb has recorded data in proton-proton collisions at $\sqrt{s} = 7$~TeV and $\sqrt{s} = 8$~TeV, with integrated luminosities of about $1\text{ fb}^{-1}$ and $2\text{ fb}^{-1}$ respectively.\footnote{LHCb has also recorded collisions of protons with lead ions, but these are not covered in this article.} The integrated luminosity has been measured using both Van der Meer scans and beam gas imaging, to accuracies of 1.7\% in $\sqrt{s}=7$~TeV collisions and 1.2\% in $\sqrt{s} = 8$~TeV collisions once measurements using the two methods are combined,\cite{Aaij:2014ida} which represent the most precise luminosity measurements at a bunched-beam hadron collider. The instantaneous luminosity at LHCb is roughly a factor of 10 lower than at ATLAS and CMS. This means that each bunch crossing typically only contains one or two proton-proton interactions. This is crucial for the identification of decays of long-lived particles, a necessity for LHCb's heavy flavor physics programme, but it also reduces the effect of pile-up interactions on detection efficiencies, and on jet reconstruction at LHCb.

\item The LHCb trigger allows events to be recorded for subsequent analysis, with the electroweak analyses requiring the presence of a candidate lepton with significant transverse momentum ($p_\text{T}$). Events containing a muon candidate with transverse momentum above 10~GeV are selected, while for electron candidates this threshold is 15~GeV. The LHCb trigger also implements a detector occupancy requirement, in order to remove a few large events which would otherwise dominate the processing time.

\item The tracking systems are located at different stages throughout the detector. Close to the interaction point is a silicon strip detector, the Vertex Locator (VELO). This is followed by a large silicon strip detector, a dipole magnet with a bending power of about 4~Tm, and three tracking stations (consisting of silicon strip detectors close to the beamline and straw drift tubes at larger angles). These provide a momentum resolution of about 1\% for particles with momentum over 100~GeV, and allow the position of the interaction point in the transverse plane to be determined to a precision of $\sim15$ microns, with the precise resolution depending on the number of tracks associated to the vertex.\cite{lhcb-det,Aaij:2014jba} 

\item Particle identification is achieved for muons using a system composed of alternating layers of iron and multiwire proportional chambers.\cite{lhcb-det} The electromagnetic calorimeter is used to identify electrons and photons. It was designed for the selection of low transverse energy ($E_\text{T}$) electromagnetic particles in the trigger. The readout is consequently optimized for the study of low $E_\text{T}$ physics, so correcting for the effects of bremsstrahlung from electrons poses a significant challenge. Regardless, electron directions, and consequently angular variables, are well determined by the tracking systems, and measurements at LHCb using electron final states provide important complementary information to studies using other final states.\cite{Aaij:2012mda} 

\item Jets are reconstructed using a particle flow algorithm, using the anti-$k_\text{T}$ algorithm,\cite{Cacciari:2008gp} and a radius parameter $R=0.5$. The LHCb jet reconstruction is optimized for jets with $p_\text{T}$ up to about 100~GeV (since a jet produced with $\eta=3$ and $p_\text{T} = 50$~GeV has total momentum over 500~GeV). The typical momentum resolution is roughly 10-15\% for jets with $p_\text{T} > 20$ GeV. The energy scale is known to an accuracy of about 3\%, determined from the accuracy with which the energy of particles in the jet are measured, and confirmed in-situ through studies of the $p_\text{T}$ balance in $Z$+jet events. Heavy flavor jets are tagged by searching for secondary vertices in within the jets. A $65\,(25)\%$ tagging efficiency is achieved for $b\,(c)$ jets with a 0.3\% light jet misidentification rate. \cite{Aaij:2013nxa,AbellanBeteta:2016ugk,Aaij:2015yqa}

\item Since LHCb is not a hermetic detector it is unsuitable for reliable determination of event-level variables such as missing energy and related quantities, and instead relies on other approaches to select, for example, events containing a $W$ boson.

\end{itemize}

\section{Studying QCD at LHCb with electroweak bosons}
\subsection{Motivation}
\begin{figure}[t]
\centerline{\includegraphics[width=0.6\textwidth]{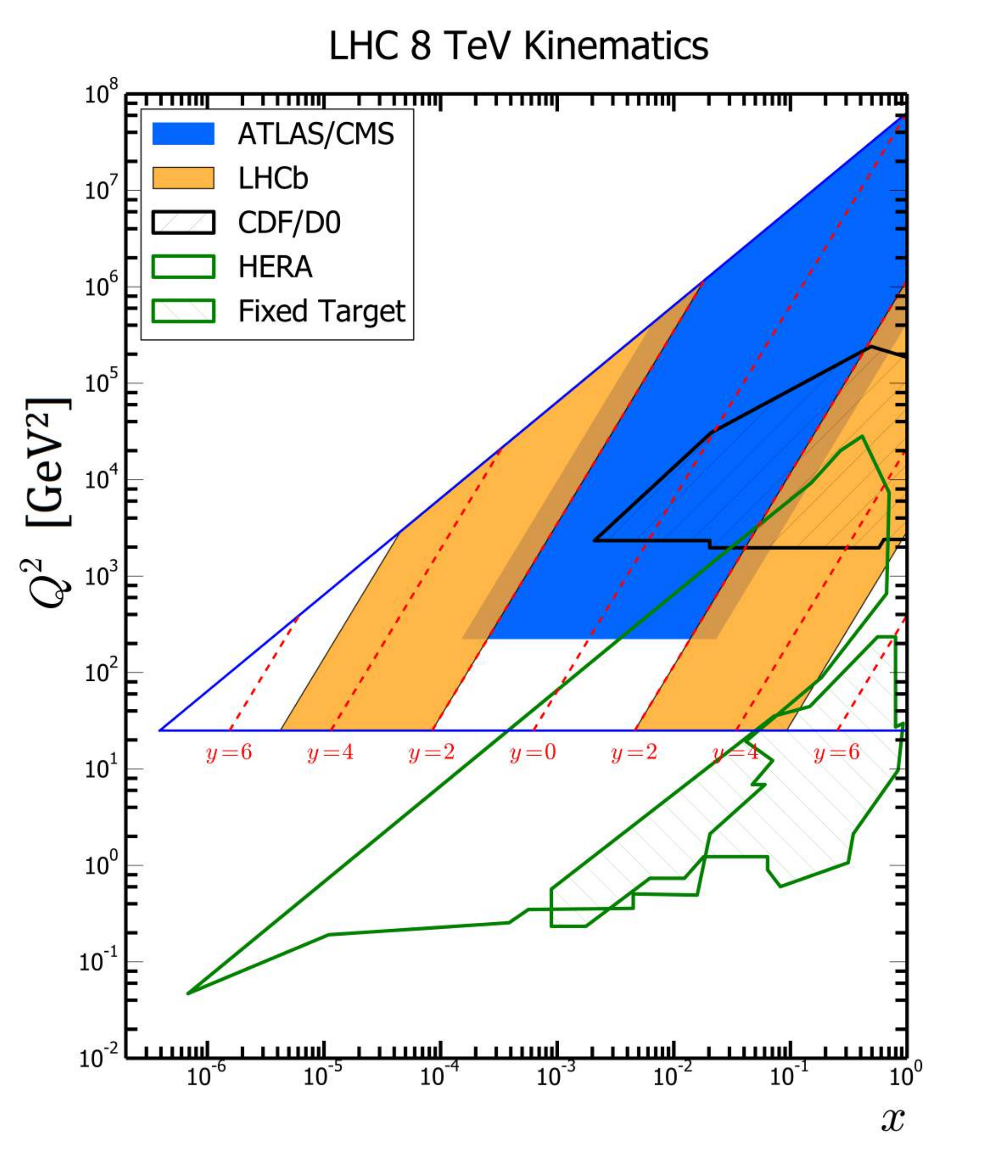}}
\vspace*{8pt}
\caption{The regions in the $x-Q^2$ plane probed at LHCb and at other LHC and previous experiments. Adapted from Ref.~\protect\refcite{Farry20162181}.\protect\label{fig1}}
\end{figure}

Partonic cross-sections are well known for $W$ and $Z$ boson production, with uncertainties at the level of 1-2\%.\cite{Martin:2009iq} The PDF uncertainties on theoretical predictions for production cross-sections in proton-proton collisions can be significantly larger. Consequently measurements of electroweak boson production cross-sections provide benchmark tests of the proton structure, and can be used to test different approaches for modeling PDFs, and to place constraints on PDF fits, reducing PDF uncertainties.\cite{Butterworth:2015oua} Differential cross-sections with respect to rapidity are expected to be particularly useful for the study of PDFs, since the rapidity ($y$) of a boson with mass $m$ can be related to the fractions of the proton longitudinal momenta carried by the colliding partons (the Bjorken-$x$ values) through

\begin{align}
  x_{1,2} \approx \frac{m}{\sqrt{s}}e^{\pm y},
  \label{pdf_rap}
\end{align}
for a proton-proton collision energy $\sqrt{s}$, and where the two colliding partons are labeled 1~and~2 respectively.\footnote{This condition is exact when the boson carries no momentum transverse to the beamline.} The unique acceptance at LHCb is of particular relevance for the study of PDFs.\cite{Thorne} Using the approximate relation of Eq.~\ref{pdf_rap}, it can be seen that bosons are only produced in the LHCb acceptance if the colliding partons have very asymmetric $x$-values (and momenta) - it is through this asymmetry that the boson is boosted into the forward region. Consequently LHCb probes different values of Bjorken-$x$ when studying electroweak boson production to ATLAS and CMS. The different regions probed by the LHC detectors in the $x-Q^2$ plane are shown in Fig.~\ref{fig1}. For electroweak boson production in the LHCb acceptance the colliding partons typically have $x$ values around $10^{-4}$ and $10^{-1}$. Both regions of $x$ probed by LHCb are of interest. Searches for physics beyond the Standard Model at much higher mass by ATLAS and CMS involve the collision of high $x$ partons, so improvements to the understanding of PDFs through LHCb measurements have a direct impact here; any constraints achieved at high $x$ by LHCb allow better understanding of Standard Model backgrounds in these searches, and will also allow any new physics signal to be understood more precisely. The low-$x$ region has not been probed directly at electroweak energy scales before. Consequently measurements at these $x$-values provide important tests of DGLAP evolution of PDFs between different energy scales.\cite{Gribov:1972ri,Dokshitzer:1977sg,ALTARELLI1977298}$^{,}$\footnote{While the PDFs are essentially non-perturbative, depending on physics around the QCD scale, their evolution between different energy scales can be modeled within pQCD using the DGLAP equations.} LHCb measurements do not just reduce the PDF uncertainties in these two regions of $x$, since the functional form of the PDFs and sum rules mean that LHCb data also have the ability to reduce the uncertainties on PDFs at intermediate values of $x$. The $W$ lepton charge asymmetry, $A^l = \frac{\sigma^+ - \sigma^-}{\sigma^+ + \sigma^-}$, measured as a function of the lepton pseudorapidity, is expected to provide important constraints on the ratio of the up quark valence PDF to the down quark valence PDF.\cite{Thorne} This is because other sources of theoretical uncertainty, and sources of experimental uncertainty, are expected to largely cancel in this ratio.

Other distributions are also of importance. The boson $p_\text{T}$ distribution is sensitive to the emission of partons, and therefore can be used to test approaches to modeling QCD radiation in different Monte Carlo simulations. The \phist variable is defined as $\phist = \tan(\phi_\text{acop.} / 2) / \cosh(\Delta\eta / 2)$, where the acoplanarity angle is defined by $\phi_\text{acop.} = \pi - |\Delta\phi|$, and where $\Delta\eta$ is the difference in pseudorapidity of the leptons produced in the boson decay and $\Delta\phi$ is the difference in azimuthal angle. This variable was first introduced for use at the Tevatron.\cite{Banfi:2010cf} It probes similar physics to that probed by the boson transverse momentum, but, being defined from angular variables, is more precisely measured. Monte Carlo event generators are not optimized using results within the LHCb acceptance, so comparisons with LHCb results provide complementary tests. The modeling of higher order radiation in pQCD is also tested by measuring differential cross-sections for the production of jets in association with electroweak bosons.

The ratio of the $W$ boson and $Z$ boson production cross-sections sees large cancellation of theoretical and many experimental uncertainties. This is also true for ratios of cross-sections of the same process measured at two different center-of-mass energies. These are predicted within the Standard Model to percent-level accuracy and better, so such measurements can constitute precision tests of the Standard Model.\cite{Thorne,Mangano:2012mh}

\subsection{Measurements at LHCb}
Measurements have been made both for inclusive final states (where any other particles can be produced with the electroweak boson), and for exclusive final states (where specific additional particles or objects are required in the final state). This section will concentrate on the inclusive measurements of $W\rightarrow\mu\nu$ and $Z\rightarrow\mu\mu$,\cite{Aaij:2014wba,Aaij:2015zlq,Aaij:2015gna} since these are the most precise measurements of electroweak boson production performed at LHCb, though measurements also exist of the $Z$ boson production cross-section in the dielectron and ditau final states.\cite{Aaij:2012mda,Aaij:2015vua,Aaij:2012bi}

Given the limited forward acceptance of LHCb, measurements are made for a fiducial region, with no extrapolation to the full phase-space. For $W$ bosons, the only fiducial requirements are that the muon produced by the $W$ boson decay has $p_\text{T} > 20$~GeV and $2.0<\eta<4.5$. In the analysis of $Z$ boson decays, these requirements are made on both muons, in addition to the requirement that the dimuon invariant mass, $m(\mu\mu)$, is in the range $60 < m(\mu\mu) < 120$~GeV. This invariant mass requirement is necessary to remove the large virtual photon contribution at lower dimuon invariant masses, though study of the Drell-Yan process at low invariant masses is itself interesting and provides important PDF constraints at lower values of $x$. For the $Z\rightarrow\mu\mu$ process the cross-sections are measured differentially as a function of the boson rapidity, the transverse momentum, and \phist. In addition the total cross-section within the fiducial region (the `fiducial' cross-section) is found.  For the $W\rightarrow\mu\nu$ process both the fiducial cross-section and the differential cross-section as a function of the lepton pseudorapidity are measured separately for each boson charge. The $W$ lepton charge asymmetry is determined from these measurements, as are ratios between the different fiducial cross-sections. These measurements are performed for both the $\sqrt{s} = 7$ and $8$~TeV datasets separately. 

The cross-sections are typically calculated from the number of events selected, the efficiencies associated with selection requirements and detecting the events, the purity of the selected sample and the integrated luminosity. Where necessary distributions are unfolded to correct for the effect of bin-to-bin migrations due to the detector response.  The inclusive measurements are corrected for the effects of final state radiation (FSR), allowing final states containing different leptons can be directly compared. These correction factors are typically determined using HERWIG++ and PYTHIA8,\cite{Bahr:2008pv,Sjostrand:2007gs} with the difference between the two setting the uncertainty associated with the correction.

The $Z\rightarrow\mu\mu$ sample is selected by searching for events containing opposite sign muon candidates that fulfill the fiducial criteria. Additional requirements are used to increase the purity of the $W\rightarrow\mu\nu$ sample: muons are required to be isolated from other activity in the event and associated to a proton-proton interaction point. For both the $W$ and $Z$ boson samples at least one muon is required to be responsible for the event passing the trigger. The purity of the selected $Z$ boson sample is over 99\% in the dimuon final state, with the largest backgrounds, arising from hadron misidentification and semileptonic decays of heavy flavor particles, determined directly from data. The purity of the $W$ boson sample is about 80\% for both $W^+$ and $W^-$, and is determined by fitting the muon $p_\text{T}$ distribution in each pseudorapidity bin, with the signal template shape determined from simulation.

\begin{figure}[t]
\centerline{\includegraphics[width=0.65\textwidth]{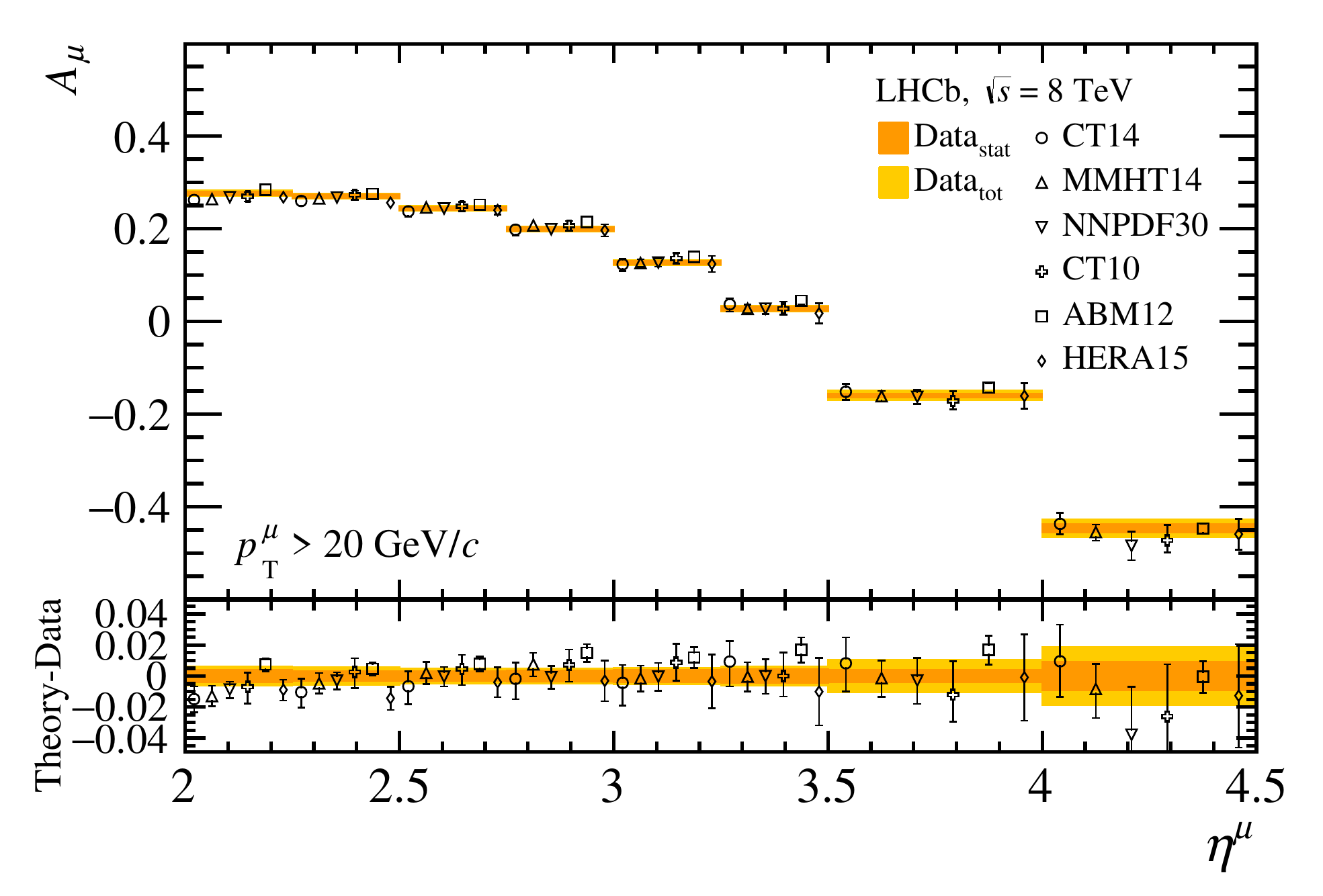}}
\vspace*{8pt}
\caption{The $W$ lepton charge asymmetry measured by LHCb in the forward region. The bands correspond to the data, with the inner band giving the statistical uncertainty, and the outer band giving the total uncertainty. The points correspond to $\mathcal{O}(\alpha_s^2)$ predictions from different PDF sets, with the uncertainties due to PDF knowledge, factorization and renormalization scale variation, and from the value of $\alpha_s$. The points are displaced horizontally in each bin for ease of comparison to data.\protect\label{fig2} From Ref.~\protect\refcite{Aaij:2015zlq}.}
\end{figure}

Detection efficiencies are typically evaluated from data using tag-and-probe techniques in $Z\rightarrow\mu\mu$ events. The muon identification efficiency is determined by only placing the muon identification requirements on one of the leptons, the tag. The fraction of events where the other lepton, the probe, is identified as a muon defines this efficiency. The method is validated by applying it to simulation, where no bias is observed. The uncertainty associated with the efficiency determination is therefore set by the size of the sample used. The efficiencies for track reconstruction and for the triggering on a single lepton are found in a similar way, save that the efficiency associated with the occupancy requirement in the trigger is found separately. This requirement has negligible inefficiency in events containing only one proton-proton interaction. Distributions for events with multiple interactions are constructed using data selected with a minimum bias trigger configuration and from collisions where only one proton-proton interaction takes place, and where an electroweak boson is produced. These distributions can then be used to determine the efficiency of the occupancy requirement. This is cross-checked using additional triggers with looser occupancy requirements; the efficiency is found to be consistent between the two methods at the per-mille level. The efficiency of the additional requirements applied to select the $W$ boson sample (that the muon is isolated and associated to a primary vertex, and that there are no other muons in the event) is determined by measuring the effect of these requirements on the high-purity $Z$ boson sample.
\begin{figure}[t]
\centerline{\includegraphics[width=0.65\textwidth]{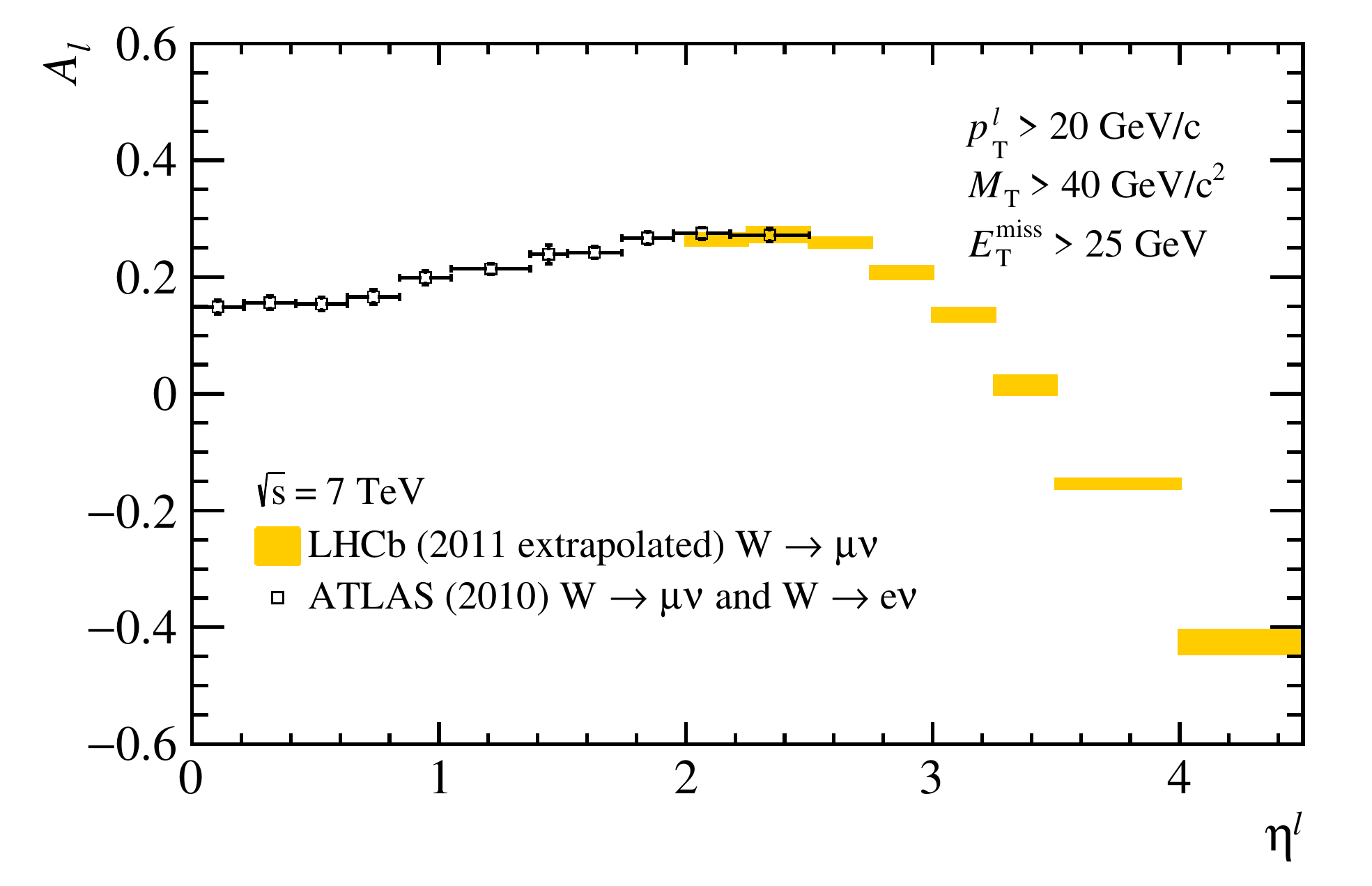}}
\vspace*{8pt}
\caption{The LHCb data for the $W$ lepton charge asymmetry are extrapolated for the additional fiducial requirements made in the ATLAS measurement, with the lepton-neutrino system required to have missing $E_\text{T} > 25$~GeV and a transverse mass above 40~GeV, and then compared to the ATLAS data. This extrapolation introduces additional uncertainties which are shown on the LHCb measurement.\label{fig3} From Ref.~\protect\refcite{Aaij:2015gna}, using methods outlined in Ref.~\protect\refcite{LHCb-CONF-2013-005}.}
\end{figure}

The integrated luminosity is known at the level of $1.7\,(1.2)\%$, for the $\sqrt{s} = 7\,(8)$~TeV measurements.~\cite{Aaij:2014ida} The next largest uncertainty arises from knowledge of the beam energy, which is known to an accuracy of 0.65\%.\cite{Wenninger:1546734} While the measurement of the cross-section is independent of the beam energy, the cross-section is quoted for a particular value of $\sqrt{s}$. Since any comparison to theory relies on this energy, an uncertainty is necessary. A shift in the beam energy of 0.65\% directly corresponds to a change in the predicted cross-section of about 1\%. This uncertainty is included on the experimental result, and should be treated as fully correlated with the same uncertainty on measurements at the other LHC experiments. The next largest uncertainties are due to knowledge of the muon reconstruction efficiencies.

The cross-sections for $W\rightarrow\mu\nu$ and $Z\rightarrow\mu\mu$ production are compared to $\mathcal{O}(\alpha_s^2)$ predictions calculated using the FEWZ generator with different PDF sets.\cite{Gavin:2010az} Theoretical uncertainties account for variations of the factorization and renormalization scales by factors of two around their nominal value of the boson mass, for PDF uncertainties, and for knowledge of the coupling strength of the strong force, $\alpha_s$. Excellent agreement is seen between the data and the theoretical predictions.

\begin{figure}[t]
\centerline{\includegraphics[width=0.65\textwidth]{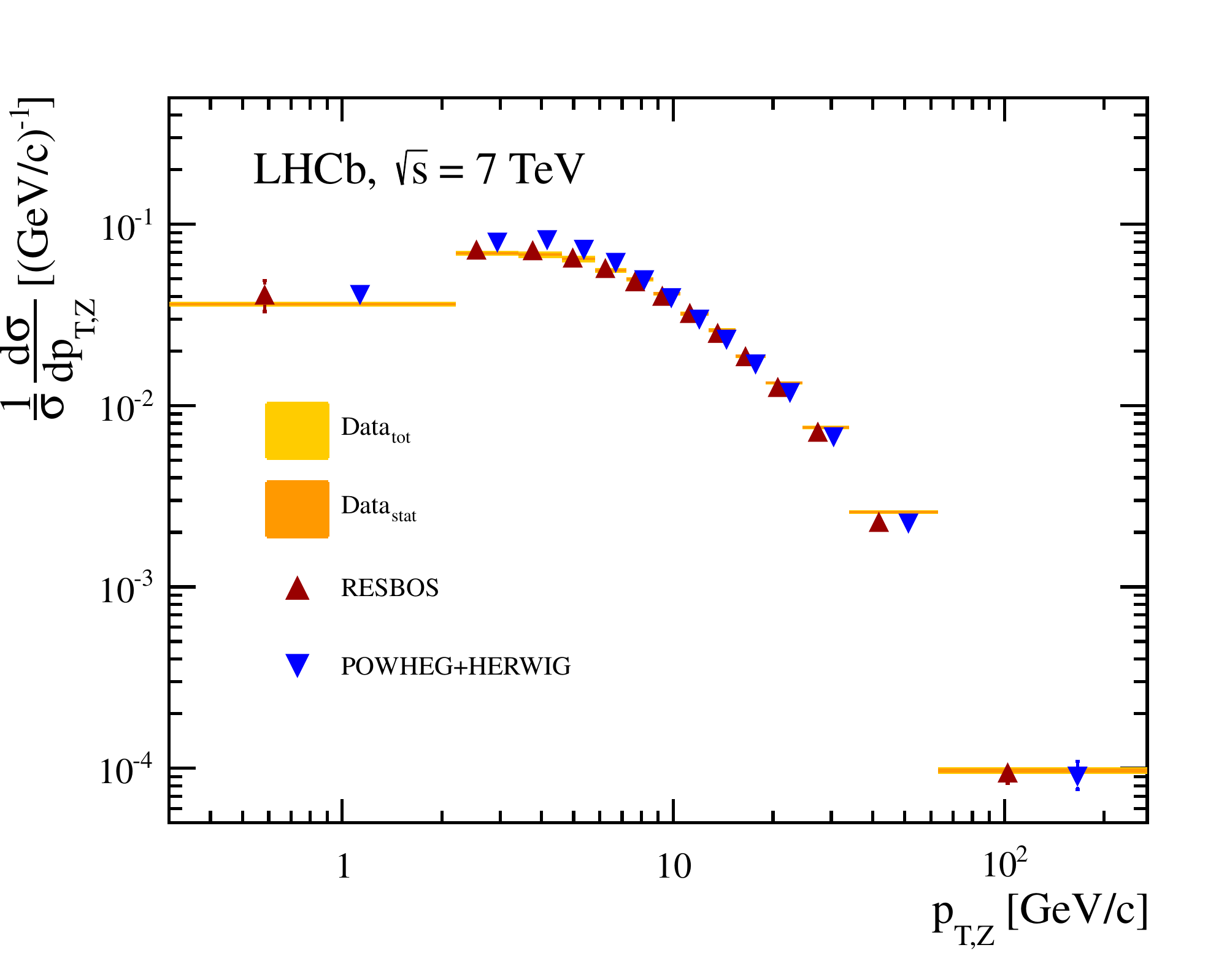}}
\vspace*{8pt}
\caption{The normalized differential cross-section as a function of the $Z$ boson $p_\text{T}$ measured at LHCb. The bands correspond to the data, with the inner band giving the statistical uncertainty, and the outer band giving the total uncertainty. The points show the theoretical predictions. The points are displaced horizontally in each bin for ease of comparison to data.\protect\label{fig4} From Ref.~\protect\refcite{Aaij:2015gna}.}
\end{figure}

Differential cross-sections are also measured, and the data are compared to the same predictions as a function of the $Z$ boson rapidity, and, for the $W$ boson, as a function of the muon pseudorapidity. The $W$ lepton charge asymmetry is also calculated and is shown for the 8~TeV measurement in Fig.~\ref{fig2}. The asymmetry changes sign at high muon pseudorapidity, as the $W^{-}$ cross-section becomes larger than the $W^{+}$ cross-section. This arises from the $V-A$ structure of the weak force: muons tend to be produced in the direction of the initial state quark, while anti-muons tend to be produced in the direction of the initial state anti-quark. The main theory uncertainty associated with the asymmetry is due to the PDFs, since other effects cancel. The high precision of the data relative to the theory uncertainties and the spread in predictions from different PDF sets gives the LHCb data considerable power to constrain the PDFs (the overall impact of LHCb measurements is discussed briefly below). The LHCb phase space is also complementary to that probed at ATLAS and CMS; Fig.~\ref{fig3} compares the ATLAS results for the differential charge asymmetry as a function of lepton pseudorapidity to those determined at LHCb for $\sqrt{s} = 7$~TeV collisions, with simulation used to extrapolate the LHCb measurement for the effects of additional fiducial requirements present in the ATLAS measurement. LHCb offers significant additional coverage to that provided by the `general purpose' LHC detectors.

\begin{figure}[t]
\centerline{\includegraphics[width=0.55\textwidth]{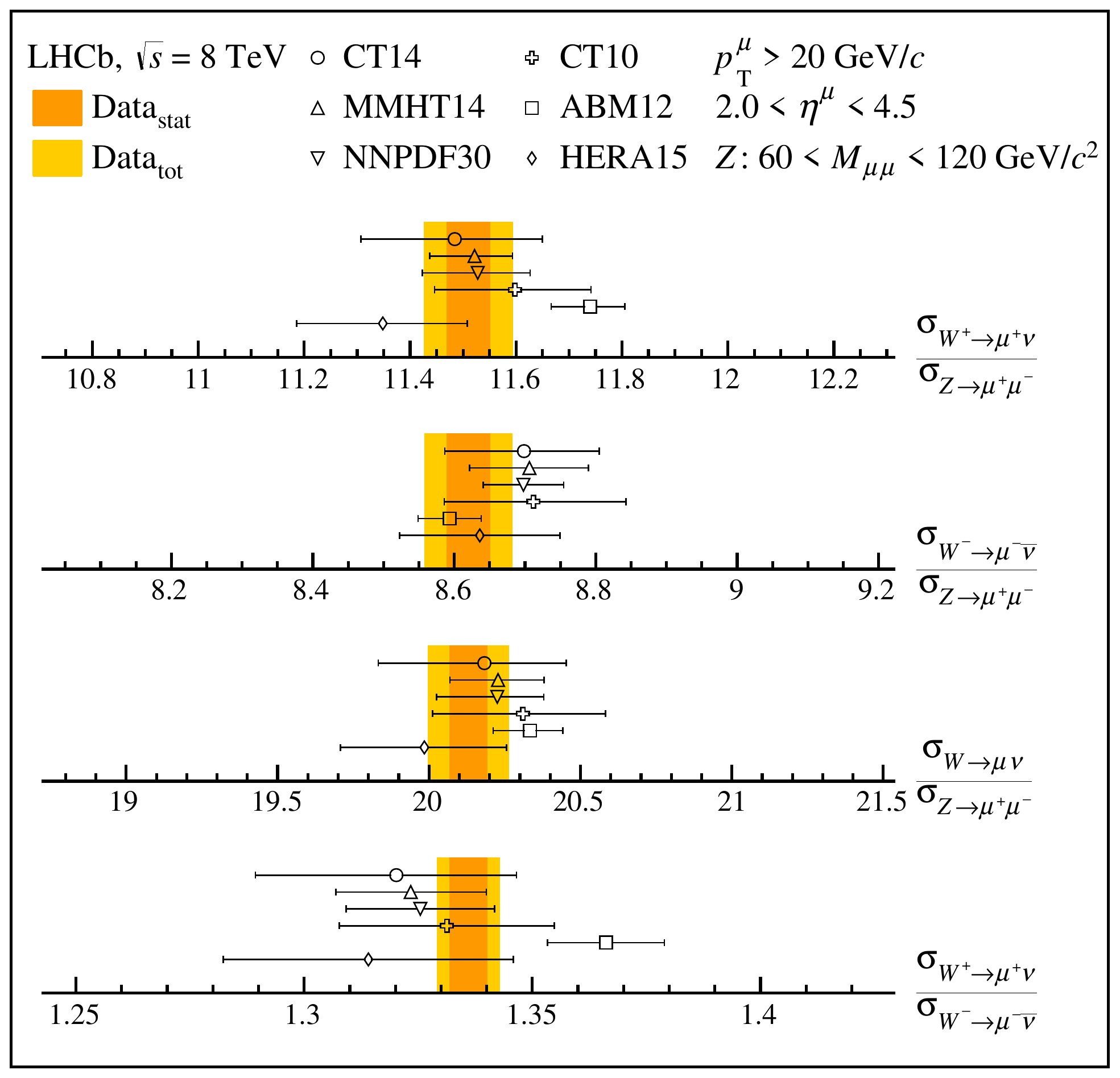}}
\vspace*{8pt}
\caption{Ratios of fiducial cross-sections at $\sqrt{s} = 8$~TeV. The bands correspond to the data, with the inner band giving the statistical uncertainty, and the outer band giving the total uncertainty. The points correspond to $\mathcal{O}(\alpha_s^2)$ predictions from different PDF sets, with the uncertainties due to PDF knowledge, factorization and renormalization scale variation, and the value of $\alpha_s$.\protect\label{fig5} From Ref.~\protect\refcite{Aaij:2015zlq}.}
\end{figure}
The differential cross-section for $Z$ boson production as a function of $p_\text{T}$ and \phist is also compared to predictions from different Monte Carlo event generators. An example is shown in Fig~\ref{fig4}, where the data collected at $\sqrt{s} = 7$~TeV are compared to predictions calculated with RESBOS and with POWHEG-BOX.\cite{Landry:2002ix,Ladinsky:1993zn,Balazs:1997xd,Alioli:2008gx} The RESBOS prediction is accurate to next-to-leading order (NLO) in $\alpha_s$, and includes the effect of soft gluon emission through the resummation of large logarithms (to next-to-next-to-leading logarithm accuracy). The POWHEG-BOX prediction is accurate to NLO, with higher order effects included through the use of a parton shower, here implemented using HERWIG.\cite{Corcella:2000bw} Reasonable agreement is seen in both cases. The same conclusions are reached through studies of the \phist variable, which probes similar physics.

\begin{figure}[t]
\centerline{\includegraphics[width=0.55\textwidth]{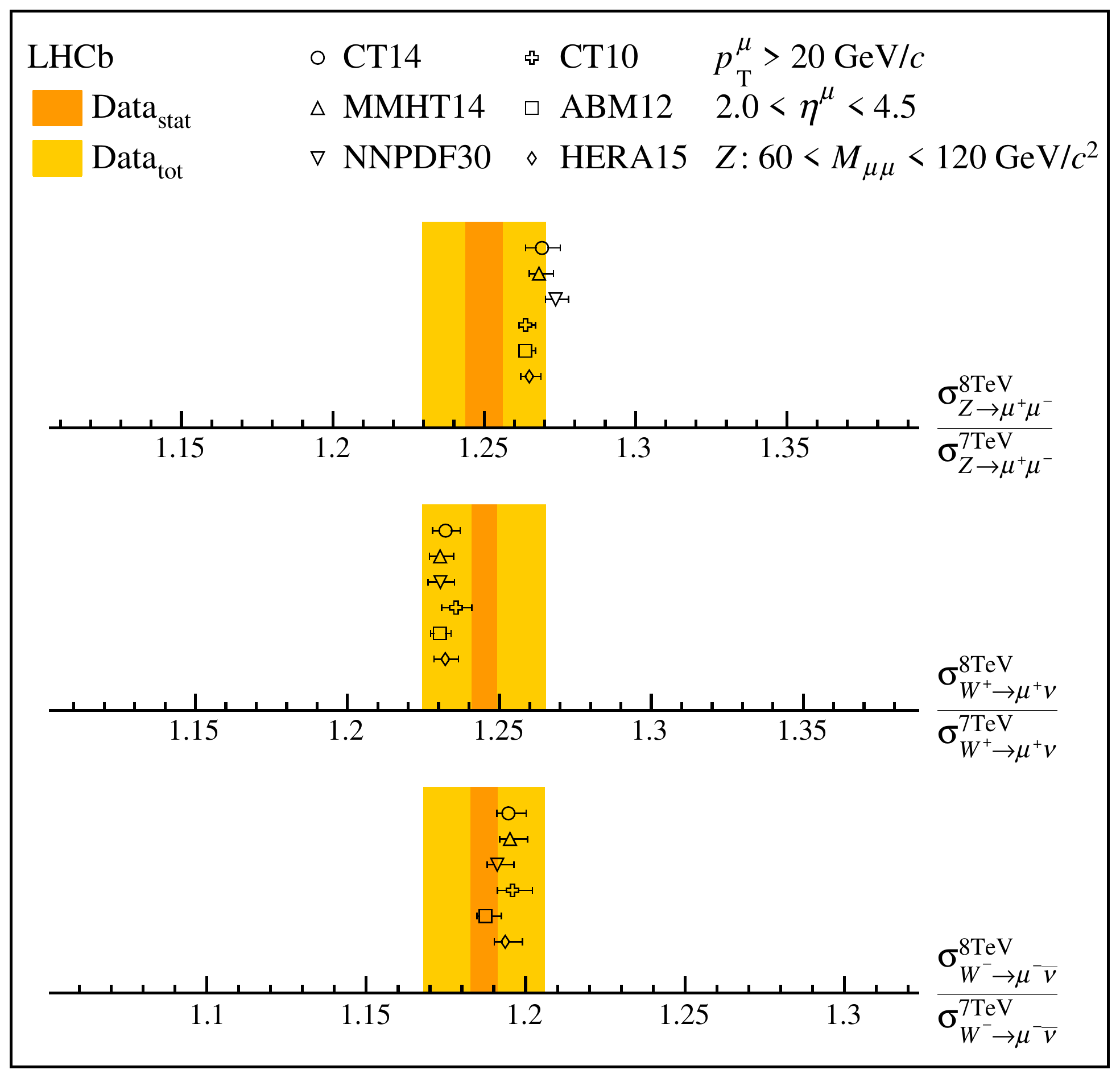}}
\vspace*{8pt}
\caption{Ratios of fiducial cross-sections at different center-of-mass energies. The bands correspond to the data, with the inner band giving the statistical uncertainty, and the outer band giving the total uncertainty. The points correspond to $\mathcal{O}(\alpha_s^2)$ predictions from different PDF sets, with the uncertainties due to PDF knowledge, factorization and renormalization scale variation, and the value of $\alpha_s$.\protect\label{fig6} From Ref.~\protect\refcite{Aaij:2015zlq}.}
\end{figure}

The ratios of fiducial cross-sections are also studied, and compared to the fixed order predictions. The results for $\sqrt{s} = 8$~TeV are shown in Fig.~\ref{fig5}; PDF uncertainties cancel considerably in the ratio of $W$ and $Z$ production cross-section, and the measurement constitutes a percent-level test of the Standard Model. All the ratio measurements are consistent with Standard Model predictions calculated at $\mathcal{O}(\alpha_s^2)$ using the FEWZ generator.\cite{Gavin:2010az} Cross-section ratios can also be formed between the two collision energies for the same process. These are shown in Fig.~\ref{fig6}. The dominant experimental uncertainty is due to the knowledge of the integrated luminosity, since the uncertainty is only partially (55\%) correlated between the two datasets.\cite{Aaij:2015zlq} The theory uncertainties largely cancel, yielding predictions of per-mille accuracy. The ratios for $W^+$, $W^-$, and $Z$ boson production cross-sections are all consistent with the Standard Model predictions. Ratios of these ratios, for example $\frac{\sigma^Z_{\text{8 TeV}}}{\sigma^Z_{\text{7 TeV}}} / \frac{\sigma^{W^{+}}_{\text{8 TeV}}}{\sigma^{W^{+}}_{\text{7 TeV}}}$, are of interest since the residual luminosity uncertainty cancels, allowing extremely precise experimental measurements to be compared to per-mille level theoretical predictions calculated at $\mathcal{O}(\alpha_s^2)$. Some of these ratios show a $\sim2\sigma$ discrepancy from the expected values within the Standard Model, so extensions of these measurements with the $\sqrt{s} = 13$~TeV dataset currently being recorded will prove interesting. Similar measurements at the other LHC experiments are also of interest.

Measurements of $Z$ boson production using the dielectron and ditau final states give results consistent with the dimuon final state measurements, probing the same physics.\cite{Aaij:2012mda,Aaij:2015vua,Aaij:2012bi}  While the experimental uncertainties are larger in the dielectron and ditau channels, many of these uncertainties are independent from those evaluated in the dimuon measurements, and consequently these results provide important additional information. These measurements also demonstrate the ability of the LHCb detector to reconstruct electrons with high transverse momenta and to distinguish final states involving taus from background.

In addition to the inclusive measurements, LHCb has reported differential measurements of the $Z+\text{jet}$ and $W+\text{jet}$ production cross-sections.\cite{Aaij:2013nxa,AbellanBeteta:2016ugk,Aaij:2015yqa} These collisions contain a greater fraction of events where the initial state contains a gluon, and consequently have more sensitivity to the gluon PDF. In addition, these measurements can be used to further constrain the down quark PDF at high-$x$.\cite{Farry:2015xha} The largest uncertainties on these results arise from the jet energy scale and from the purity of the $W+\text{jet}$ sample. The measurements are consistent with the Standard Model. Ratios of these cross-sections are also measured; the ratio of the $W^++\text{jet}$ fiducial cross-section to the $W^-+\text{jet}$ fiducial cross-section is the most precise of these measurements, determined to an accuracy of about 3.4\%, and is in good agreement with Standard Model predictions. Measurements have also been made of production in association with heavy flavor particles.\cite{Aaij:2014hea,Aaij:2014gta,Aaij:2015cha} These can be used to test the need for 4- and 5-flavor PDF schemes, and to constrain the strange and charm content of the proton. The jet measurements also serve as key inputs to understanding important backgrounds to top production at LHCb, which was observed in the LHCb dataset and reported in Ref.~\refcite{top}.

\subsection{Impact and potential for future studies}
The LHCb results have been included in PDF fits with significant impact,\cite{Harland-Lang:2014zoa,Ball:2014uwa,Dulat:2015mca,Alekhin:2013nda} reducing the uncertainty on PDFs by as much as 50\% in some regions of Bjorken-$x$.\cite{ubiali} 
Future measurements will continue to have an important impact.
The inclusive cross-sections and their ratios will also be measured again at $\sqrt{s}=13$~TeV, which provides new constraints on PDFs, with the sensitivity to PDFs at low Bjorken-$x$ increasing. This programme of study has already commenced in Ref.~\refcite{Aaij:2016mgv}.
It is also clear from Fig.~\ref{fig1} that measurements of the Drell-Yan process at lower masses will also increase sensitivity to lower values of Bjorken-$x$. This is not just of interest for PDF constraints; QCD at low-$x$ is an important field of study in its own right.\cite{d'Enterria:2006nb}

\section{Tests of electroweak physics at LHCb}
\label{sect:s2tw}
\subsection{Measuring the weak mixing angle}
The differential cross-section for the decay of the $Z/\gamma^*$ to a dilepton final state can be written as
\begin{align}
  \begin{split}
\frac {\rd^2 \sigma } {\rd\cos\theta^{*}\rd\phi^{*}}
\propto &\Bigl[(1+\cos^2\theta^{*}) +A_0 \frac{1}{2}(1-3\cos^2\theta^{*}) \\&+ A_1\sin2\theta^{*}\cos\phi^{*} + A_2\frac{1}{2}\sin^2\theta^{*}\cos2\phi^{*}\\&
  +A_3\sin\theta^{*}\cos\phi^{*} + A_4\cos\theta^{*} + A_5 \sin^2 \theta^* \sin2\phi^*\\&+A_6\sin2\theta^*\sin{\phi^*} + A_7\sin{\theta^*}\sin{\phi^*}\Bigr],
  \end{split}
  \label{eq:angcoeff}
\end{align}
where $\theta^*$ and $\phi^*$ are the polar and azimuthal angles of the negatively charged lepton in the Collins-Soper frame.\cite{Khachatryan:2015paa,PhysRevD.16.2219} This is the dilepton rest-frame, with the $z$-axis bisecting the direction of the two incoming protons. The use of this frame reduces effects from boson transverse momentum. The angular coefficients $A_i$ contain information about the polarization of the boson produced, but also depend on the boson couplings to fermions. The coefficient $A_4$ introduces a forward-backward asymmetry in $\theta^*$, arising from the presence of both vector and axial-vector couplings. This asymmetry can be defined simply as $A_\text{FB} = \frac{N(\cos\theta^*>0) - N(\cos\theta^*<0)}{N(\cos\theta^*>0) + N(\cos\theta^*<0)}$, where $N$ is the number of events fulfilling the relevant condition. This asymmetry is enhanced by interference of the $Z$ boson with the virtual photon; the asymmetry is larger when the dilepton invariant mass lies away from the $Z$ pole mass. The asymmetry exhibits significant dependence on the dilepton mass, taking a different sign at high mass and at low mass. Since the asymmetry depends directly on the vector and axial-vector couplings, it is sensitive to the weak mixing angle which relates the two. Specifically, such measurements can be used to determine the effective leptonic weak mixing angle, \sttw, which differs from $\sin^2(\theta_\text{W})$ through higher order effects.

Measurements of \sttw have been made at previous colliders, and the current world average is dominated by the combination of measurements at LEP and at SLD, which give $\sttw = 0.23153\pm0.00016$.\cite{ALEPH:2005ab} LHC measurements are not expected to rival this precision. However, it remains interesting to measure this variable, because the two most precise determinations of the angle (one at LEP, the other at SLD) differ by over 3 standard deviations.\cite{ALEPH:2005ab,PhysRevLett.84.5945} Apparent process dependence in determining \sttw from different results should be investigated with further measurements, since new physics may become visible in such studies and, even if new physics is not forthcoming, hadron collider measurements may favor one of the precision measurements.

Measurements of $A_\text{FB}$ at the LHC show the most sensitivity to \sttw when made in the forward region. At a rapidity of 0, the initial state is symmetric, with the initial state quark as likely to come from either proton, so no forward-backward asymmetry is generated; there is complete dilution of the parton-level asymmetry in the proton-proton collisions. It is for this reason that there is little sensitivity to \sttw at low rapidities. However, at larger rapidities, the dilution is lessened. $Z$ bosons produced in the LHCb acceptance are produced by the collision of one parton at high $x$ and one at low $x$. The PDFs dictate that the high $x$ parton tends to be a valence quark, with the low $x$ parton being the antiquark. Consequently, a large forward-backward asymmetry is visible in the proton collisions, providing greater sensitivity to \sttw. In addition, the dilution between the parton-level asymmetry and that arising in proton collisions is predicted more precisely in the forward region. This reduces the impact of the PDF uncertainty on any extraction of \sttw at LHCb.

\begin{figure}[t]
  \begin{center}
  \includegraphics[width=0.475\textwidth]{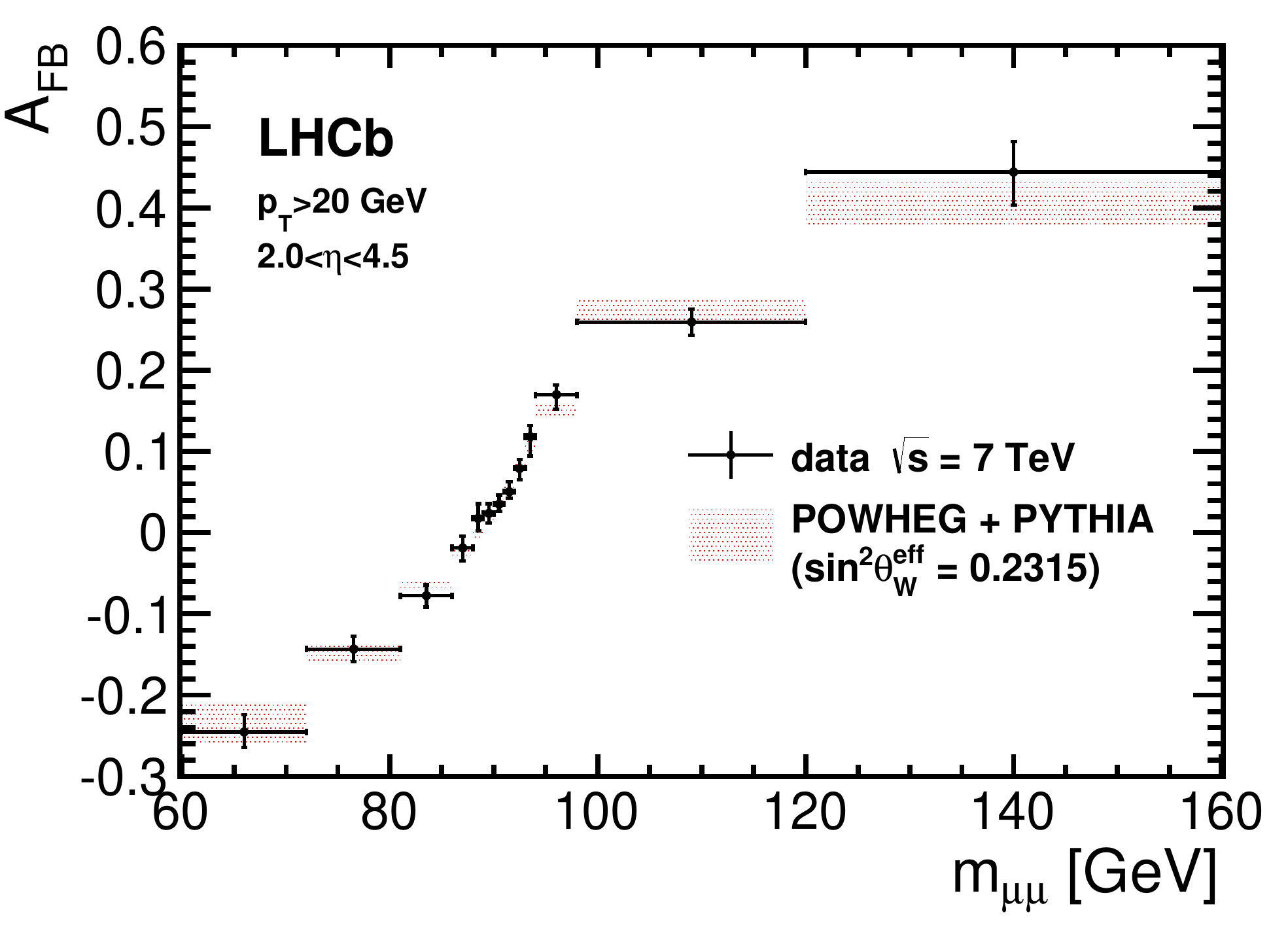}
  \includegraphics[width=0.475\textwidth]{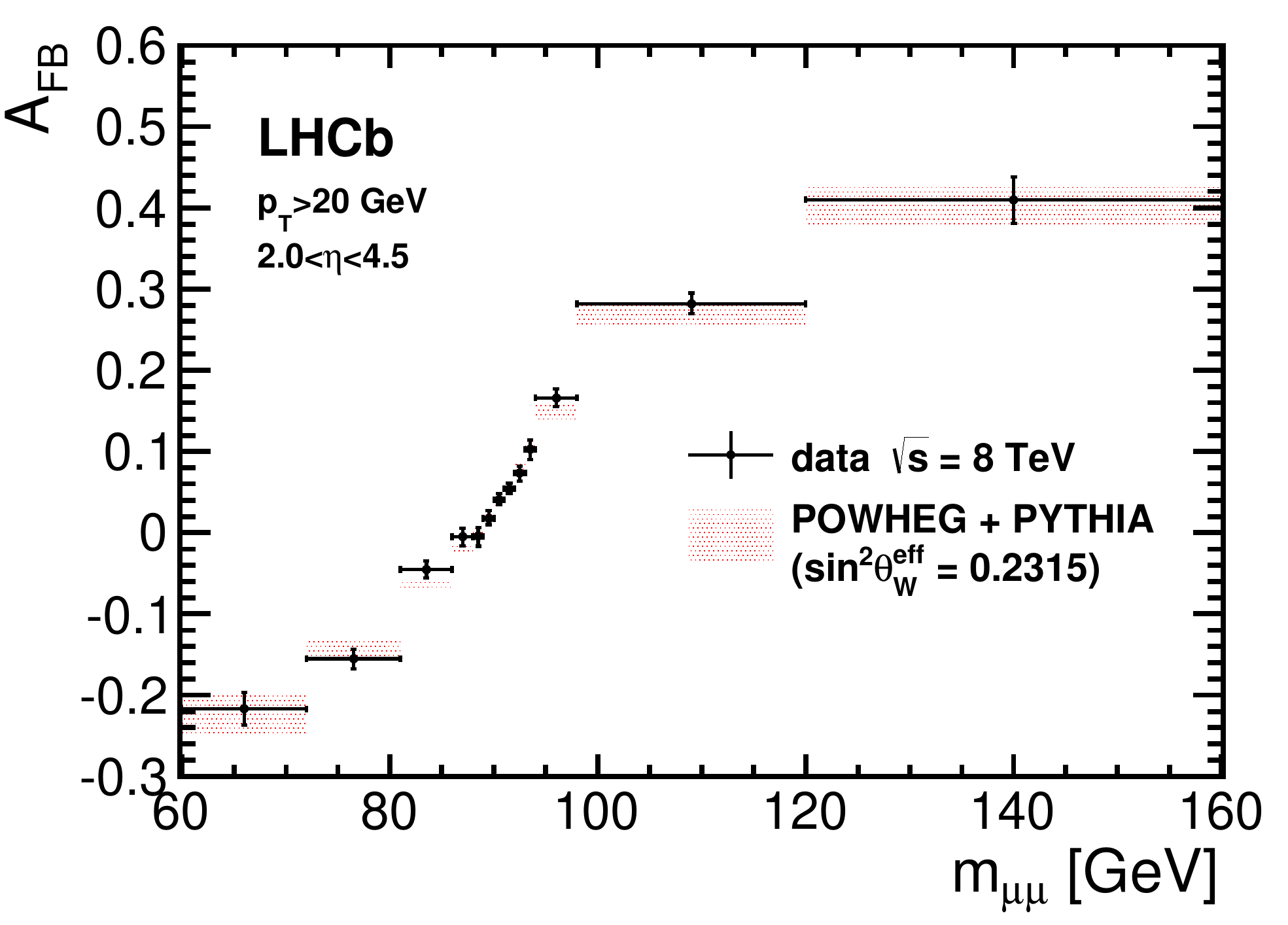}
  \end{center}
\vspace*{8pt}
\caption{The forward-backward asymmetry in $\sqrt{s} = 7$~TeV collisions (left) and in $\sqrt{s} = 7$~TeV collisions (right), measured as a function of the dimuon invariant mass. The points show the data, while the bands are theoretical predictions generated using POWHEG-BOX and PYTHIA8 for the world-average value of \sttw. The theory uncertainties arise from PDFs, scale variation, the value of $\alpha_s$, and from modeling of FSR. \protect\label{fig7} From Ref.~\protect\refcite{Aaij:2015lka}.}
\end{figure}

\begin{figure}[t]
\centerline{\includegraphics[width=1.0\textwidth]{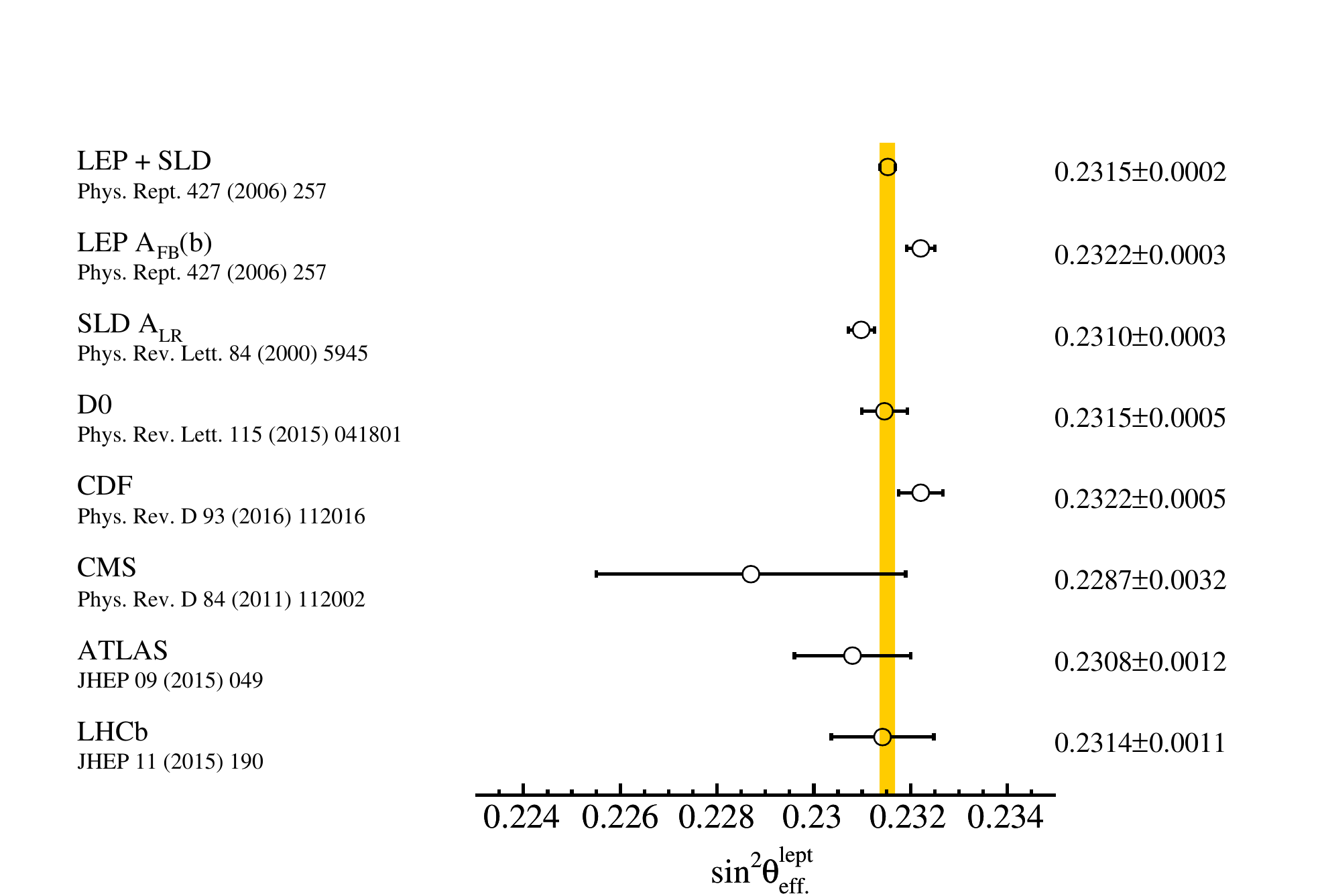}}
\vspace*{8pt}
\caption{Different measurements of \sttw. The band shows the LEP and SLD combination. Also shown are the most precise determinations of \sttw at the two experiments, and measurements at the Tevatron and the LHC. While no update of the CMS result has been published, CMS have made precise determinations of the angular coefficients set out in Eq.~\ref{eq:angcoeff}, which probe related physics, in Ref.~\protect\refcite{Khachatryan:2015paa}.\protect\label{fig8}}
\end{figure}

LHCb has measured the forward-backward asymmetry as a function of the dilepton invariant mass.\cite{Aaij:2015lka} The dimuon final state is considered since this provides the best experimental precision at LHCb. The fiducial acceptance and selection of events is the same as in the inclusive cross-section analysis, save that the upper limit on the invariant mass is relaxed to 160~GeV, since the large asymmetry at high dimuon invariant masses provides additional sensitivity to \sttw despite the limited event yield. Selection efficiencies and purities are also determined using the same techniques as in the inclusive measurement. The analysis is performed separately for the 7~TeV and 8~TeV datasets, with the extracted values of \sttw then combined into a single result. The asymmetry is unfolded to correct for detector effects and for bin-to-bin migrations in the dimuon invariant mass. No correction is applied for the dilution between the parton-level asymmetry and that measured in proton-proton collisions; instead the theoretical predictions are made at the particle level. Various sources of uncertainty are considered in the asymmetry measurement. Those associated with background levels and efficiency are negligible. Significant uncertainties instead arise from knowledge of the detector alignment (and consequently the momentum scale) and from modeling the detector resolution of the dimuon invariant mass and its effects. The asymmetry is shown in Fig.~\ref{fig7}. The value of \sttw is determined through a series of fits using templates of the asymmetry generated with different values of \sttw. These templates are produced using POWHEG-BOX with the parton shower effects simulated using PYTHIA8, and using NNPDF2.3 PDFs.\cite{Alioli:2008gx,Sjostrand:2007gs,Ball:2012cx} The best value of \sttw is determined as $\sttw = 0.23142\pm0.00073\pm0.00052\pm0.00056$, where the first uncertainty is statistical, the second is due to experimental systematic effects and the third is due to theoretical uncertainties in determining \sttw from the forward-backward asymmetry. The largest theory uncertainty arises from the PDF knowledge. This is determined from the NNPDF2.3 uncertainty band,\cite{Ball:2012cx} which is larger than the change in result from switching the PDF set to CT10.\cite{Lai:2010vv} Smaller theoretical uncertainties arise from the choice of the factorization and renormalization scales, from the value of $\alpha_s$ and from knowledge of the effects of FSR. The result is found to be robust against changes in the choice of event generator used to provide templates, and is currently the most precise determination of \sttw at the LHC. It is compared to other measurements in Fig.~\ref{fig8}.


\subsection{Potential for future studies}
The LHCb measurement of the weak mixing angle is currently statistically limited, so further studies using the $\sqrt{s} = 13$~TeV dataset currently being recorded are expected to reduce the uncertainty. The limiting theoretical uncertainty is due to parton distribution functions. However, this may also be reduced in future measurements with the use of future PDF sets which will use new LHC data to reduce PDF uncertainties, or through in situ constraints placed on PDFs using the LHCb $Z\rightarrow\mu\mu$ dataset, or through binning the measurement in bins of rapidity, giving more weight to the regions most sensitive to \sttw. Such an approach will be considered when the statistical uncertainty no longer dominates.

Aside from measurements of the weak mixing angle, it is clear that LHCb can make a key contribution in at least one other important area of electroweak physics: the measurement of the $W$ boson mass.\cite{Bozzi:2015zja} Such a measurement can be made at LHCb by considering the $p_\text{T}$ of the charged lepton produced in the $W$ boson decay, since this distribution is sensitive to the $W$ boson mass. The largest uncertainties associated with a $W$ boson mass measurement at any LHC experiment are expected to arise from knowledge of the PDFs. However, the PDF uncertainties are anti-correlated between the ATLAS and CMS acceptance and the LHCb acceptance, such that any measurement at LHCb has a significant impact on the overall uncertainty in any combined LHC measurement. It is claimed that a combination of a measurement in the central region with a LHCb measurement will therefore be more precise than a combination of two measurements in the central region, and that a combination of ATLAS, CMS and LHCb measurements should be a factor of 1.3 more precise than a combination of ATLAS and CMS alone.\cite{Bozzi:2015zja} A combined measurement of the $W$ boson mass using information from ATLAS, CMS and LHCb could reach a precision of about 10~MeV, improving on the current world average (15~MeV), assuming that the production processes are well understood.\cite{Bozzi:2015zja} A measurement of the $W$ boson mass will therefore prove an important extension of LHCb's study of electroweak physics.

\section{Conclusions}
LHCb measurements of electroweak boson production and decay are important contributions to the study of Standard Model phenomenology at the LHC. Measurements of electroweak boson production cross-sections are consistent with Standard Model predictions, and have provided important constraints on PDFs at both low-$x$ and at high-$x$, and have significantly reduced PDF uncertainties when included in global PDF fits. Various ratios of fiducial cross-sections are also able to test the Standard Model below the percent-level. Such studies will be complemented by similar measurements using the higher energy data currently being recorded. LHCb is also able to probe the weak mixing angle, through the study of the forward-backward asymmetry in $Z$ boson decays. The forward region is particularly sensitive to such a measurement, since the dilution between the parton-level and proton-level asymmetry is reduced. The LHCb measurement is the most precise at the LHC, and is still statistically limited. Improved precision in this measurement is expected with analysis of the data currently being recorded.

\section*{Acknowledgments}
I am thankful to members of the LHCb collaboration for discussions: S.~Farry, P.~Ilten, T.~Shears, M.~Vesterinen, V.~Vagnoni, and D.R.~Ward.
%This section should come before the References. Dedications and funding
%information may also be included here.

\bibliographystyle{ws-mpla}
\bibliography{review}

\end{document}